\documentclass[conference]{IEEEtran}
\IEEEoverridecommandlockouts
\usepackage{cite}
\usepackage[inkscapelatex=false]{svg}
\usepackage{amsmath,amssymb,amsfonts}
\usepackage{algorithmic}
\usepackage{graphicx}
\usepackage{textcomp}
\usepackage{xcolor}
\usepackage{array}      % (optional) for better column definitions like >{\centering\arraybackslash}p{...}
\usepackage{caption}    % (optional) for caption spacing control
\usepackage{booktabs}   % for \toprule, \midrule, \bottomrule
\usepackage{siunitx}    % for \SI and units
\usepackage{multirow}
\usepackage[table,xcdraw]{xcolor}
\usepackage{colortbl}

\usepackage[linesnumbered,ruled,vlined]{algorithm2e}
\SetKw{KwRet}{return}  % ensures \KwRet exists
\SetCommentSty{scriptsize}

\def\BibTeX{{\rm B\kern-.05em{\sc i\kern-.025em b}\kern-.08em
    T\kern-.1667em\lower.7ex\hbox{E}\kern-.125emX}}
\begin{document}

% \title{Performance Evaluation of GitOps Tools in Intent-Based Networking Orchestration Scenario}
\title{Performance Evaluation of Intent-Based Networking Scenarios: A GitOps and Nephio Approach}

\author{\IEEEauthorblockN{Saptarshi Ghosh,
Ioannis Mavromatis, Konstantinos Antonakoglou, and Konstantinos Katsaros\\
\IEEEauthorblockA{Digital Catapult, United Kingdom\\}
Emails: \{saptarshi.ghosh, ioannis.mavromatis, konstantinos.antonakoglou, kostas.katsaros\}@digicatapult.org.uk}}

\maketitle

\begin{abstract}

GitOps has emerged as a foundational paradigm for managing cloud-native infrastructures by enabling declarative configuration, version-controlled state, and automated reconciliation between intents and runtime deployments. Despite its widespread adoption, the performance and scalability of GitOps tools in Intent-Based Networking (IBN) scenarios are insufficiently evaluated. This paper presents a reproducible, metric-driven benchmarking, assessing the latency and resource overheads of three widely used GitOps operators: Argo CD, Flux CD, and ConfigSync. We conduct controlled experiments under both single- and multi-intent scenarios, capturing key performance indicators such as latency and resource consumption. Our results highlight trade-offs between the tools in terms of determinism, resource efficiency, and responsiveness. We further investigate a realistic orchestration scenario, using Nephio as our orchestrator, to quantify the processing latency and overhead in declarative end-to-end deployment pipelines. Our findings can offer valuable insights for tool selection and optimisation in future autonomous network orchestration systems.
\end{abstract}


\section{Introduction}
\IEEEPARstart{T}{he} increasing complexity of modern cloud-native, Intent-Based Networking (IBN) \cite{in1} infrastructure has stimulated a shift from imperative to declarative deployment models. GitOps \cite{in2} has emerged as a compelling operational paradigm that uses Git as the Single Source of Truth (SSoT) for managing declarative infrastructure and application configurations. By continually synchronising the runtime states of deployments in Kubernetes (K8s) clusters with their desired states stored in the associated Git repositories through automated agents (i.e., Reconciliation Operators), GitOps enhances auditability, traceability, and overall system reliability.

In the context of cloud-native IBN, network orchestrators automate the deployment of desired network states, including all necessary assets, such as Virtual Network Functions (VNFs) and their interconnections, from a declarative manifest. For this paper, we use Nephio \cite{in6} as our desired orchestrator. Nephio captures an ``Intent'' as configuration values defined through Kubernetes Resource Model (KRM), decoupled from the application code and declarative deployment manifests, a concept referred to as Configuration as Data (CaD) \cite{in6}. GitOps provides an automated Continuous Deployment (CD) pipeline by tracking the desired states in a Git repository, comparing them with the corresponding runtime states of the orchestrated VNFs, and synchronising the states as soon as it detects any drift between them. When GitOps and CaD are combined, we have an IBN orchestrator capable of translating customer intent into a set of desired states, encoded in declarative manifests, and pushing it to the Git repository for deployment. To that extent, in CaD-enabled frameworks like Nephio, the terms ``intent" \& ``desired-state" become synonymous. In such a framework, all Key Performance Indicators (KPIs) (e.g., reconciler's latency, resource utilisation, etc.) contribute to the End-to-End (E2E) performance of an IBN Orchestration \cite{in4,in5}.

This paper builds upon the above principles and presents a systematic benchmarking of a cloud-native IBN orchestrator, broken down into two parts, i.e., the GitOps and the CaD pipelines. We first design a reproducible benchmarking pipeline using Kubernetes, GitLab \cite{gitlab} and three leading GitOps tools (ArgoCD \cite{argo}, FluxCD \cite{flux}, and ConfigSync \cite{csync}) to simulate realistic GitOps workflows. Later, we evaluate a CaD pipeline based on Nephio. We capture a set of latency and resource utilisation metrics, and provide an empirical evaluation under various scaling conditions. Our work aims to offer a quantitative performance baseline informing practitioners and researchers about the strengths and limitations of GitOps-enabled IBN orchestrators.

The remainder of this paper is organised as follows: Sec. \ref{sec:soa} outlines related works, Sec. \ref{sec:sys-des} describes the system design of the benchmarking system and its integration with Nephio. Sec. \ref{sec:result} presents the experimental results and their analysis, and summarises the findings. We conclude in Sec. \ref{sec:conclusion} summarising the contribution and future outlook of this work.

\section{State-of-the-Art \& Motivation}
\label{sec:soa}
This work builds upon our previous work \cite{camino}, that demonstrates a bespoke IBN Orchestration platform, named Cloud-native Autonomous Management and Intent-based Orchestrator (CAMINO), utilising Nephio and Kubernetes, where we experienced certain limitations of ConfigSync. In this section, we first summarise the state of the art in GitOps and CaD, as well as their utilisation in cloud-native use cases, followed by a gap analysis that motivates this study.

The authors in \cite{rw4} recommend using GitOps over DevOps, demonstrating the advantages of a pull-based, declarative deployment approach over its push-based counterpart through configuration changes and rollbacks using Argo CD. The study \cite{rw2} presents a comprehensive guide to implementing Flux CD with a Kubernetes cluster. However, it lacks any quantitative results. The work \cite{rw3} demonstrates a proof-of-concept using tools included in the Cloud Native Computing Foundation (CNCF) landscape, which validates the feasibility of GitOps in an IoT Edge-computing solution utilising Argo CD. However, it could not provide any remarks regarding scaling due to the limited number of worker nodes in the cluster. The analysis in \cite{rw5} compares the computational overhead between GitOps and DevOps, indicating a trade-off between efficiency and cost. It shows a $17\%$ reduction in total lines of code, accompanied by a $53\%$ decrease in the average number of lines of code per file. However, there is a $75\%$ increase in the number of files, along with an $80\%$ increase in the depth of the directory tree. The methodology proposed in \cite{rw6} demonstrates an Argo-CD-based Containerised Network Function (CNF) configuration framework using YANG Models and NETCONF, with scenarios for full and partial configuration changes. The implementation described in \cite{rw8} is closest to our work, which utilises Nephio for IBN on a KIND-based K8s cluster on OpenStack and deploys a fixed Private 5G topology by integrating the Argo CD operator, replacing ConfigSync. Although it provides quantitative results on resource utilisation, it lacks in measuring latency. Finally, the evaluation in \cite{rw1} compares the performance of reconciliation with that of traditional configuration management using Argo CD and Ansible \cite{ansible_docs}, respectively. The performance analysis reveals that Argo CD consistently outperforms Ansible in handling misconfiguration changes, network configuration drift, and dependency updates.

Based on the ground study above, we identify two research gaps that we intend to address. First, the lack of a quantitative performance comparison based on latency and resource utilisation of the GitOps tools; and second, a comparison between single or simultaneous intent deployments (i.e., state changes), investigating how the reconciliation operators perform under different scaling conditions. The primary motivation for this work is to enhance CAMINO's performance by introducing a more capable reconciler. Building a bespoke experimental pipeline, we intend to evaluate the different frameworks in detail, identify an alternative to ConfigSync as the reconciler for CAMINO, and share our findings with the research community for further considerations.

\begin{table}[t]
\renewcommand{\arraystretch}{0.72}
\caption{List of Notations Used}
\centering
\footnotesize
    \begin{tabular}{lp{0.68\columnwidth}}
        \toprule
        \textbf{Notation} & \textbf{Description} \\
        \midrule
        $EP$     & Experiment Parameters $\{p,m,r,c\}$ \\
        $p \in [\mathrm{a}\text{-}\mathrm{z}]^+$      & Prefix String to identify GitOps tool \\
        $m \in \mathbb{N}$      & Maximum number of deployments or replicas \\
        $r \in \mathbb{N}$    & Number of representation \\
        $c \in \mathbb{N}$      & Step count or range $[1:m:c]$ \\
        $S_{\text{template}}$  & Template of Desired State Manifest: YAML File \\
        $R_{\text{template}}$  & Template of Reconciler Manifest: YAML File \\
        $S_{\text{desired}}$   & Desired State of a deployment: YAML File \\
        $S_{\text{runtime}}$   & Runtime State of a deployment: YAML File \\
        $O_{\text{recon}}$     & Reconciler Operator Instance \\
        $\Delta S$      & State Drift: \texttt{Diff}($S_{\text{desired}}$,$S_{\text{runtime}}$) \\
        $t_{\text{push}} \in \mathbb{Q}$  & Time to Push (in Sec.)   \\
        $t_{\text{recon}} \in \mathbb{Q}$ & Time to Reconcile (in Sec.) \\
        $t_{\text{deploy}} \in \mathbb{Q}$ & Time to Deploy (in Sec.) \\
        $t_{\text{healthy}} \in \mathbb{Q}$ & Time to reach Healthy status (in Sec.) \\
        $t_{\text{hydrate}} \in \mathbb{Q}$ & Time to Hydrate a Dry Kpt package (in Sec.) \\
        $t_{\text{inproc}} \in \mathbb{Q}$ & Nephio's Intent Processing Time \\
        $u_{\text{cpu}} \in \mathbb{Q}$ & CPU Utilization by active reconciler in Millicore \\
        $u_{\text{mem}} \in \mathbb{Q}$ & Memory Utilization by active reconciler in MiB \\
        $K_{\text{attr}}$ & KPI Attributes \\
        $K_{\text{attr}}^{(m,r,c)}$ & Aggregated KPI Attributes per experiment \\
        $[l:u:c]$ & Discrete range: $l$ to $u$ (inclusive) with increment $c$ \\
        \bottomrule
    \end{tabular}\label{tab:acronyms}
\end{table}

\section{System Design}
\label{sec:sys-des}
This section outlines the high-level system design of the benchmarking methodology, highlighting the algorithms employed and integration with the Nephio platform. Table \ref{tab:acronyms} lists the various notations used in the following text.

\subsection{Benchmarking System}
\begin{figure}[t]
    \centering
    \includegraphics[width=1\linewidth]{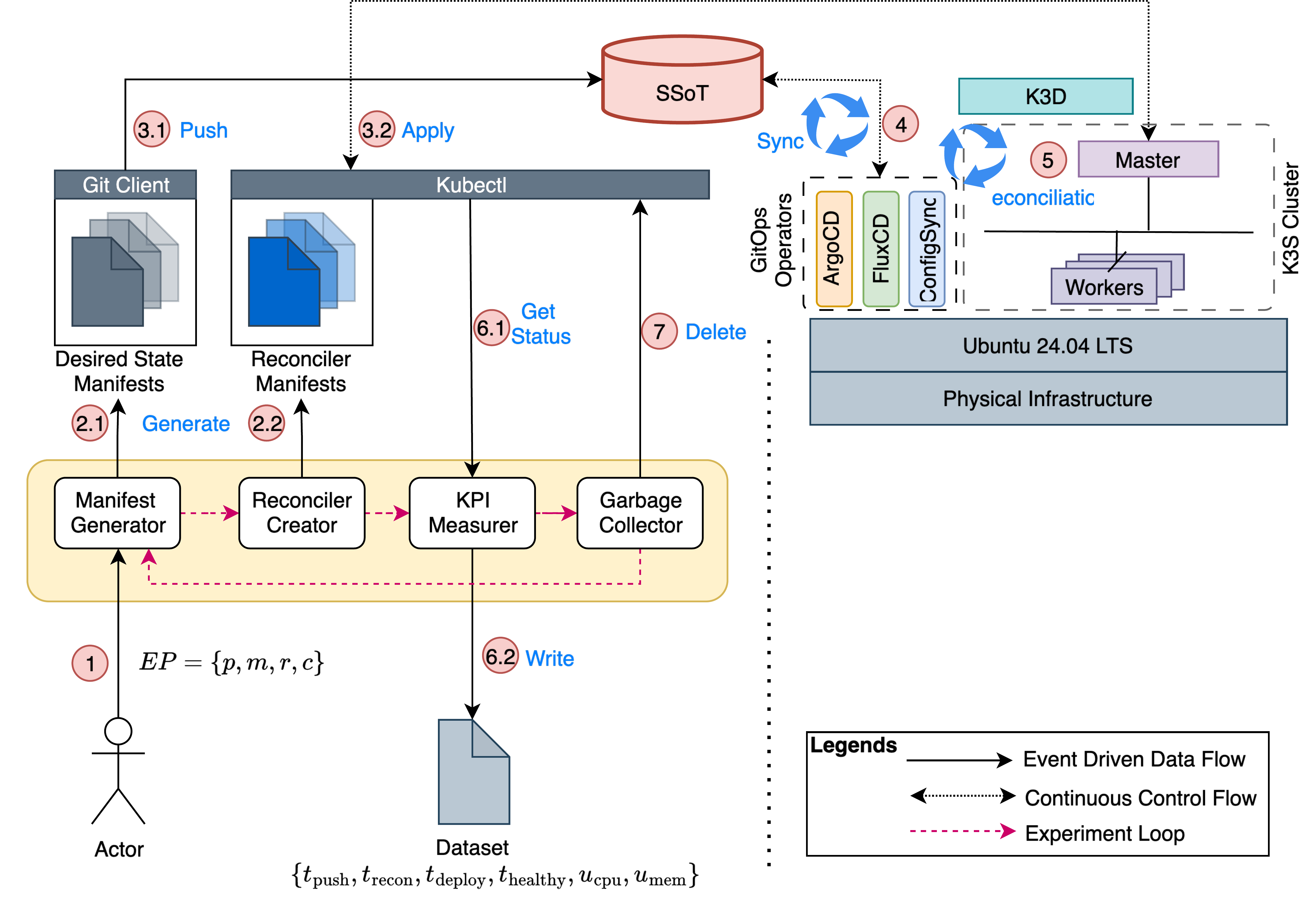}
    \caption{Benchmarking pipeline for our GitOps investigation}
    \label{fig:bm-infra}
\end{figure}

\begin{table}[t]
\renewcommand{\arraystretch}{0.72}
\caption{Testbed Specifications}
\centering
\footnotesize
    \begin{tabular}{ll}
        \toprule
        \textbf{Resource} & \textbf{Specification} \\
        \midrule
        CPU & 2 Sockets, 4 Cores each,  clocked at 4.2 GHz   \\
        RAM & 16 GB DDR4 \\
        Kubernetes Cluster & K3s with K3d wrapper  \\
        \bottomrule
    \end{tabular}
    \label{tab:specs}
\end{table}

Fig.~\ref{fig:bm-infra} illustrates the schematic diagram of our GitOps benchmarking pipeline. It comprises a benchmarking system, a Git repository serving as the SSoT, and the Kubernetes (K8s) cluster coexisting with three GitOps operators: Argo CD, Flux CD, and ConfigSync. Table \ref{tab:specs} summarises the testbed specification. The control plane of each GitOps tool is isolated within its respective namespace, which hosts the reconciliation operators and other control plane components. We measure the resource consumption of these namespaces to compare the GitOps tools in isolation. The lifecycle of an intent is divided into four stages (Generation, Synchronisation, Reconciliation, and Deployment stages), while the benchmarking experiments loop in four phases (Manifest Generation, Reconciler Generation, KPI Measurement, and Garbage Collection).

The Actor initiates an experiment through the Manifest Generation stage by providing a set of parameters $EP=\{p, m, r, c\}$, where $p$ denotes a string (prefix) to select the GitOps tools to use and name related variables, $m$ is the maximum number of application manifest (apps) or replicas to deploy, $r$ represents the number of repetitions of the same experiment, and $c$ is the step count of the range $[1:m]$. The benchmarking system runs as a four-phase experimentation loop. In Phase 1, the \textit{Manifest Generator} generates a set of manifest files based on $EP$ from a state template $S_{\text{template}}$ representing a collection of Desired States $\{S_{\text{desired}}\}_{i=1}^{n}$ where $n$ is the number of deployments (multi-app) or replicas (single-app). Therefore, the experimentation loop is set to run for $|\{S_{\text{desired}}\}|=\lceil{\frac{mr}{c}}\rceil$ iterations (Alg.~\ref{algo:man-gen}). Phase 1 concludes by invoking a Git Push operation that uploads the desired states into the Git repository (SSoT) with a specific directory structure. We measure the time to push as $t_{\text{{push}}}$. In Phase 2, the \textit{Reconciler Generator} generates a set of reconciler object manifests $\{O_{\text{recon}}\}_{i=1}^{n}$ corresponding to $S_{\text{desired}}^{i}$ derived from a reconciler template $R_{\text{template}}^{(p)}$, depending on the GitOps tools identified by the prefix $p$ (Alg.~\ref{alg:recon-gen}).

In the implementation, instead of creating separate repositories for each GitOps operator to hook, we create a single repository containing multiple directories, each representing an $S_{\text{desired}}^{i}$ and specified their unique directory path in the corresponding $O_{\text{recon}}^{i}$ with a shared Git URL. Applying the manifests $\{O_{\text{recon}}^{i} | i \in [1:m:c]\}$ establishes Webhooks dynamically between the $S_{\text{desired}}^{i}$ at SSoT and the $O_{\text{recon}}^{i}$ operators to track them. The GitOps operators compare the state drift $\Delta S$ as the drift between the tracked desired states $S_{\text{desired}}^{i}$ and their corresponding runtime states $S_{\text{runtime}}^{i}$. Although realising $\Delta S$ numerically is challenging, for convenience we shall denote $\Delta S = 0$ if $S_\text{desired}^{i}=S_{\text{runtime}}^{i} \forall i$ and $\Delta S \ne 0$ otherwise. In our implementation, we determine $\Delta S$ by comparing the revision string of the repository. We establish a Webhook between the SSoT and the GitOps operators authenticated by a Personal Access Token (PAT) to keep the desired state in sync. In Stages 2 and 3 of the workflow, the GitOps operator synchronises the $\{S_{\text{desired}}\}_{i=1}^{n}$ as the associated Webhook detects a Push operation, followed by calculating $\Delta S$. We measure the synchronisation and reconciliation time taken by the operators as $t_{\text{sync}}$ and $t_{\text{recon}}$, respectively.

In Stage 4 of the E2E workflow, Kubernetes manages deployments of the desired states with a drift. We measure the time to deploy as $t_{\text{deploy}}$ and the time it takes for deployment to become healthy as $t_{\text{healthy}}$. In Phase 3 of the experiment loop, our \textit{Results Collector} fetches the respective namespaces of the GitOps operators, various status messages and metrics (CPU and Memory consumption) from the Kubernetes master, namely, $K_{\text{attr}}=\{t_{\text{push}}, t_{\text{sync}}, t_{\text{recon}}, t_{\text{deploy}}, t_{\text{healthy}}, u_{\text{cpu}}, u_{\text{mem}}\}$ (Alg.~\ref{alg:measure}). Finally, after measurements collection, the benchmarking system releases all the resources by clearing all associated deployments, K8s CRDs and namespaces before looping back to Phase 1.

The remainder of this section outlines each of the benchmarking phases. We performed two types of experiments for each GitOps tool: first, a single-app deployment with multiple replicas, and second, a multiple-app deployment with single replicas each, where the single-app benchmarking measures the GitOps tools' reaction time to $\Delta S$, the multi-app benchmarking measures the ability to handle simultaneous $\Delta S$ with resource consumption.

\subsubsection{Phase 1: Manifest Generation}
The manifest generation algorithm (Alg. \ref{algo:man-gen}) takes three inputs $(n,d_{\text{target}},p) \, | \, \forall n \in [1:m:c]$, and $d_{target}$ is the name of the sub-directory(s) Alg. \ref{algo:man-gen} creates to segregate application manifests. We use Helm to dynamically generate manifests with the app name, target namespace $NS$, and label derived from $p$.

\begin{algorithm}[t]
\DontPrintSemicolon
\caption{Generate Kubernetes manifests}\label{algo:man-gen}
\KwIn{$n$, $p$, $d_{\text{target}}$}
\KwOut{$\{S_{\text{desired}}^1, \dots, S_{\text{desired}}^n\}$}

\For{$i \gets 1$ \KwTo $n$}{
  $name \gets$ ``'\texttt{\{$p$\}-app-\{$i$\}}'' \tcp*[f]{app name}\;
  $ns \gets$ ``\texttt{\{$p$\}-ns-\{$i$\}}'' \tcp*[f]{namespace}\;
  $label \gets$ ``\texttt{\{$p$\}-label-\{$i$\}}'' \tcp*[f]{label}\;
  \texttt{makeDir}(``\{$d_{\text{target}}$\}/\{$name$\}'')\;
  $S_{\text{desired}}^i \gets S_{\text{template}}(name, ns, label)$  \tcp*[f]{Generate $S_{\text{desired}}^{i}$ manifest using HELM Template $S_{\text{template}}$}\;
  \texttt{store}($S_{\text{desired}}^i$) \tcp*[f]{write to manifest}\;
}
\KwRet{$\{S_{\text{desired}}^i\}_{i=1}^{n}$}\;
\end{algorithm}

\subsubsection{Phase 2: Reconciler Generation}
The reconciler generator algorithm (Alg. \ref{alg:recon-gen}) generates $\{O_{\text{recon}}^{i}\}_{i=1}^{n}$ of length $n=1$ for single-app and $n$ for multi-app benchmarks. However, the manifest definition varies based on the GitOps operator to target. We use operator-specific reconciler templates $R_{\text{template}}^{(p)}~|~p \in \{\texttt{"argo"}, \texttt{"flux"},\texttt{"csync"}\}$ and Helm Charts to generate  $\{O_{\text{recon}}^{i}\}_{i=1}^{n}$ dynamically, linking with the appropriate Git repository with a PAT.

\begin{algorithm}[t]
\DontPrintSemicolon
\caption{Generate reconciler manifests}\label{alg:recon-gen}
\KwIn{$n$, $p$}
\KwOut{$\{O_{\text{recon}}^1, \dots, O_{\text{recon}}^n\}$}

\For{$i = 1$ \KwTo $n$}{
  $rec$ $\gets$ ``\texttt{\{p\}-rec-\{i\}}'' \tcp*[f]{reconciler name}\;
  $ns$ $\gets$ ``\texttt{\{p\}-ns-\{i\}}'' \tcp*[f]{namespace}\;
  $url$ $\gets$ args[\texttt{cluster-url}] \tcp*[f]{K8S cluster}\;
  $br$ $\gets$ args[\texttt{git-branch}] \tcp*[f]{git branch to watch}\;
  $rdir$ $\gets$ args[\texttt{repo-dir}] \tcp*[f]{sub-dir of the $br$}\;
  \If{$p \in \{\texttt{argo}, \texttt{csync}, \texttt{flux}\}$}{
    $O_{\text{recon}}^i$ $\gets$ $R_{\text{template}}^{(p)}$ $(rec, ns, url, br, rdir)$ \tcp*[f]{Generate $O_{\text{recon}}^{i}$ manifest using HELM Template $R_{\text{template}}^{(p)}$ }\;
    \texttt{store($O_{\text{recon}}^i$)} \tcp*[f]{store manifest file}\;
  }
}
\Return ${\{O_{\text{recon}}^i\}_{i=1}^{n}}$\;
\end{algorithm}

\subsubsection{Phase 3: KPI Measurement}
The performance-measuring algorithm (Alg.~\ref{alg:measure}) is identical across all GitOps tools. We maintained this homogeneity to keep the comparison agnostic of the GitOps tools, despite some tools, e.g., ArgoCD, providing APIs to fetch the measurements. We leverage the K8s status logs fetched from Kubectl to extract attributes in $K_\mathrm{attr}$. For an experiment defined by parameters $EP$, Alg. \ref{alg:measure} measures the KPI attributes $K_\mathrm{attr}^{(m,r,c)}$ as Eq. 1.

\begin{align}
K_{\text{attr}}^{(m,r,c)} = \Big\{ \frac{1}{r} \sum_{j=1}^{r} \Big( & \sum_{i=1}^{k} t_{\text{push}}, \sum_{i=1}^{k} t_{\text{sync}},  \sum_{i=1}^{k} t_{\text{recon}}, \nonumber \\
& \sum_{i=1}^{k} t_{\text{deploy}}, \sum_{i=1}^{k} t_{\text{healthy}}, \\
& \frac{1}{k} \Big( \sum_{i=1}^{k} u_{\text{cpu}},\sum_{i=1}^{k} u_{\text{mem}} \Big) \Big)\Big\} \nonumber \\
& \; | \; \forall k \in [1 : m : c] \nonumber
\end{align}

We measure the trend of the attributes in $K_{\text{attr}}^{(m,r,c)}$ with respect to $m$ to identify a correlation between them.

\begin{algorithm}[t]
\DontPrintSemicolon
\caption{Measuring performance  metrics}
\KwOut{$K_{\text{attr}}$}\label{alg:measure}

\ForEach{$(\text{app}, \text{ns}) \in \{(\text{app}, \text{ns})\}$}{
  $t_{\text{start}} \gets$ \texttt{tstamp()} \tcp*[f]{timestamp}\;
  git.push($S_{\text{desired}}$[app]) \tcp*[f]{push desired state to repo}\;
  $t_{\text{push}} \gets \texttt{tstamp()} - t_{\text{start}}$ \tcp*[f]{calculate $t_{\text{push}}$}\;

  $t_{\text{start}} \gets$ \texttt{tstamp()} \tcp*[f]{timer reset}\;

  \While{$\Delta S \neq 0$}{
    continue\;    \tcp*[f]{drift: git revision update}
  }
  $t_{\text{sync}} \gets \texttt{tstamp()} - t_{\text{start}}$ \tcp*[f]{calculate $t_{\text{sync}}$}\;

  $t_{\text{start}} \gets$ \texttt{tstamp()} \tcp*[f]{timer reset}\;
  \While{true}{
    $t_{\text{create}} \gets$ \texttt{app.CreationTimestamp}\;

    \If{$t_{\text{create}}(\text{ns}) \neq \texttt{Null}$}{
      \textbf{break}\;
    }
  }
  $t_{\text{deploy}} \gets t_{\text{create}} - t_{\text{start}}$ \tcp*[f]{calculate $t_{\text{deploy}}$}\;

  $t_{\text{start}} \gets$ \texttt{tstamp()} \tcp*[f]{timer reset}\;

  \While{true}{
    $n_{\text{ready}} \gets$ \texttt{app.availableReplicas}\;
    $n_{\text{desired}} \gets$ \texttt{app.replicas}\;
    \If{$n_{\text{ready}} = n_{\text{desired}}$ \textbf{and} $n_{\text{desired}} \neq n_{\text{desired}}$}{
      \textbf{break}\;
    }
  } \tcp*[f]{wait until all deployed states becomes healthy}\;
  $t_{\text{healthy}} \gets$ \texttt{tstamp()} $-$ $t_{\text{start}}$\;

  $u_{\text{cpu}}, u_{\text{mem}} \gets$ \texttt{Top}($O_{\text{recon}}^{(p)},ns$ ) \tcp*[f]{Measure resource utilization using Top tool}\;
}
$K_\text{attr} \gets \{t_\text{push}, t_\text{sync}, t_\text{deploy}, t_\text{healthy}, u_\text{cpu}, u_\text{mem}\}$\;
\KwRet{$K_{\text{attr}}$}\;
\end{algorithm}

\subsubsection{Phase 4: Garbage Collection}
After every iteration of the experiment loop, the garbage collection phase runs, communicating with the K8s cluster through Kubectl to release all occupied resources by deleting any associated K8s objects (e.g., CRDs, Deployments, and Namespaces). This process eliminates the risk of miscalculation due to progressive loading by resetting the cluster after each iteration.

Our four-phase experiment loop iterates through the three different tools, considering all possible combinations of replicates and applications, and repeats $r$ times for each experiment to ensure the statistical validity of our results. Finally, all results are saved in a CSV file for further data analysis.

\begin{figure}[t]
    \centering
    \includegraphics[width=1\linewidth]{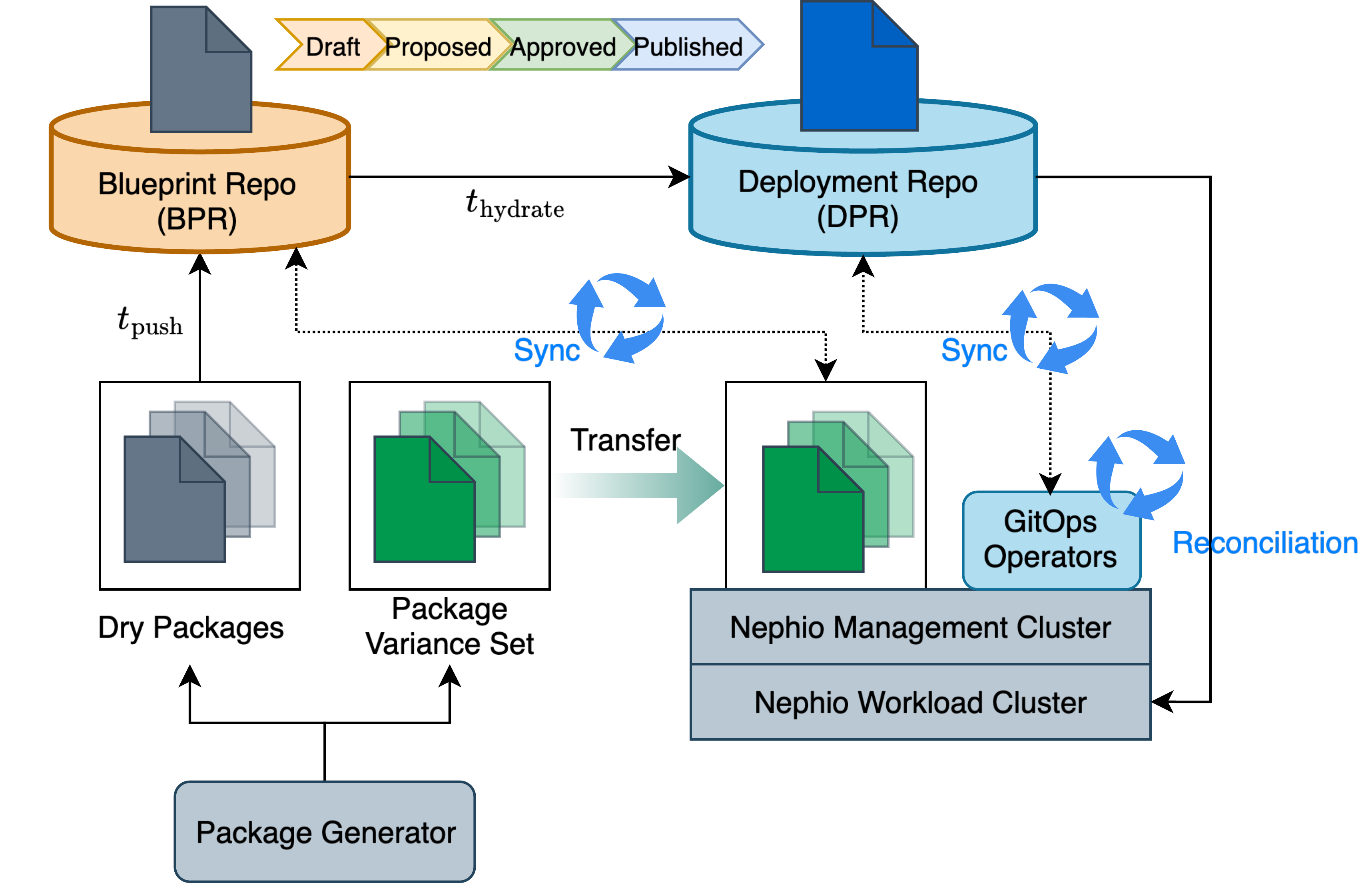}
    \caption{Nephio Integration with Benchmarking System}
    \label{fig:nephio-integration}
\end{figure}

\subsection{Nephio Integration}
Nephio's intent processing mechanism involves two repositories, namely Blueprint (BPR) and Deployment (DPR). BPR holds the intents, i.e., the set of $S_{\text{desired}}$ as Kpt packages (also called \textit{Dry Packages}). A Package Variant Set (PVS) tracks the revisions of Dry Packages at the BPR through Webhooks. Nephio leverages Package Orchestrator (Porch) to process Dry Packages (``hydrate'' them), inserting configuration data into the templates and turning them into \textit{Hydrated Packages}. During the Hydration Process, Nephio handles the Git operations, creating a branch named ``Draft" in the DPR and merging it with the main branch after processing. During the Hydration process, a package migrates through four sequential states, i.e., draft, proposed, approved, and published. Finally, published packages reside in the DPR's main branch, which the operators from $O_{\text{recon}}$ track; therefore, the integration with the Benchmarking system sets the DPR as the SSoT, as described in the previous section, introducing an additional latency for Hydration ($t_{\text{hydrate}}$). Fig. \ref{fig:nephio-integration} depicts the integration workflow.

\section{Result \& Analysis}
\label{sec:result}
This section covers the findings by analysing the experimental results of deploying a lightweight Nginx application template as $S_{\text{template}}$. We cover three test-case scenarios, first, single app deployment with scaling it by its replica with a range of $[1:100:10]$ with $r=20$ and measuring $t_{\text{push}}$, $t_{\text{sync}}$, $t_{\text{recon}}$, $t_{\text{deploy}}$; second, simultaneous multiple app deployment with single replica with a range of $[1:90:10]$ with $r=20$, measuring $t_{\text{recon}},t_{\text{deploy}}, t_{\text{healthy}}, u_{\text{cpu}}, u_{\text{mem}}$; third the latency introduced by Nephio processing single and multiple Intents. For all the experiments, we have grouped the aggregated measurements ($K_{attr}^{(m,r,c)}$) by their respective reconciler prefix $p$ and traced their corresponding trajectory of median trend lines to establish our conclusion. For multi-app deployment, we have kept $m=90$ as the default maximum pod count of K3s clusters, which is 110 to accommodate additional reconcilers' control-plane pods.

\subsection {Test-Case 1: Single-App Deployment Latency Comparison}
\begin{figure}
    \centering
    \includegraphics[width=1\linewidth]{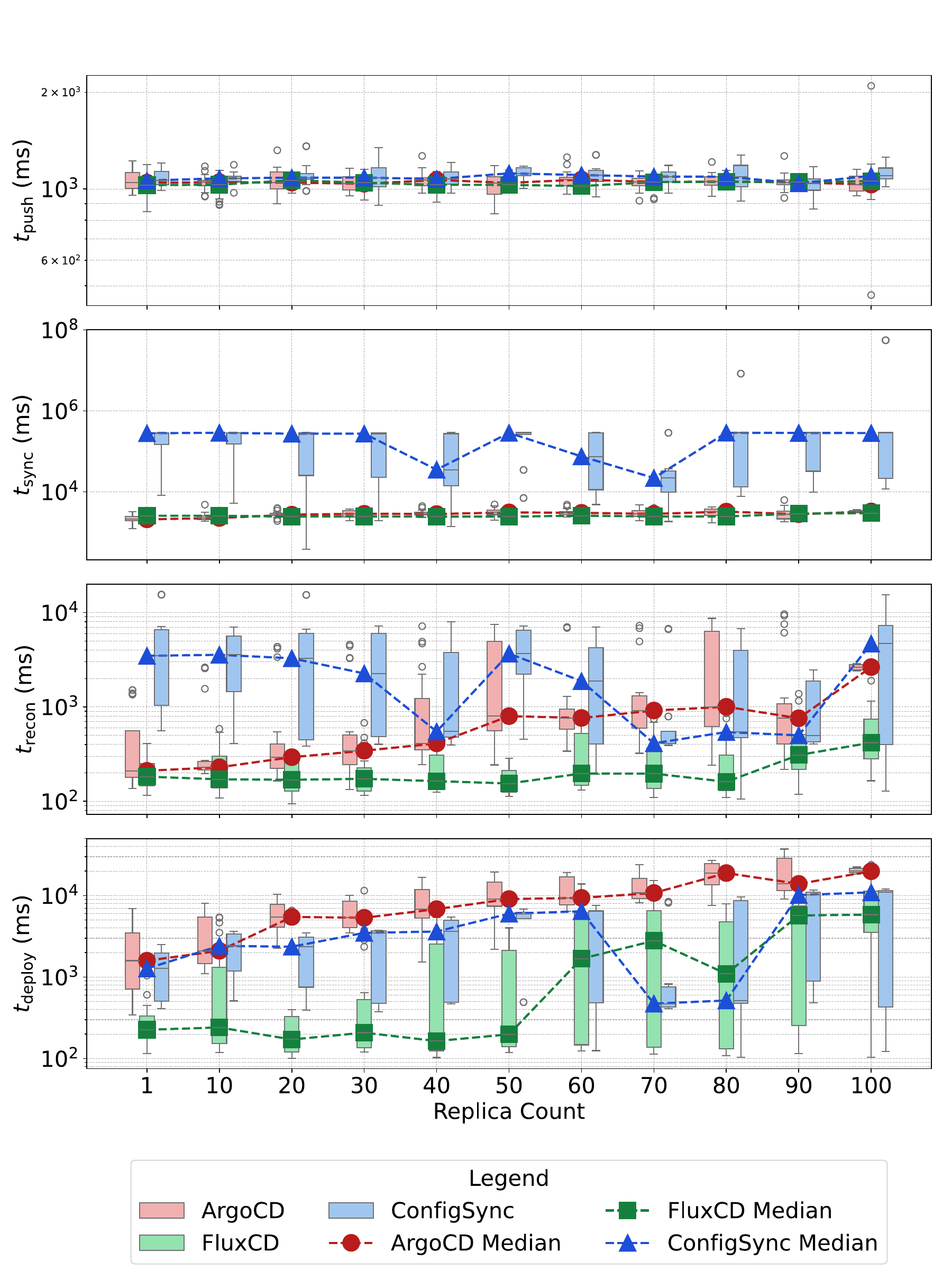}
    \caption{Latency Comparison of GitOps Tools for Single App Deployment with Scaling Replica Count}
    \label{fig:single-app}
\end{figure}

Fig. \ref{fig:single-app} depicts the combined measurements of the experiments. All three reconcilers show an identical $t_{\text{push}}$ of around \SI{1}{\second}, with outliers lying in a close neighbourhood, which is expected since $t_{\text{push}}$ is independent of the reconciliation process; however, this verifies the correctness of the tailing Push mechanism of the manifest generation. Argo CD and Flux CD outperform ConfigSync in $t_{\text{sync}}$ with identical latency bounded between \SIrange[range-phrase=--,range-units=single]{1}{10}{\second}, whereas ConfigSync took between \SIrange[range-phrase=--,range-units=single]{50}{100}{\second}. Examining the reason, we discovered that the \textit{period} attribute of RootSync (i.e., ConfigSync's reconciler) is not persistent on the open-source version of ConfigSync if it runs outside of Google Cloud. Comparison of $t_{\text{recon}}$ shows Argo CD being the fastest (\SIrange[range-phrase=--,range-units=single]{0.5}{0.8}{\second}), followed by Flux CD (\SIrange[range-phrase=--,range-units=single]{0.5}{5}{\second}) both showing a predictable trajectory. However, ConfigSync shows significant fluctuation between \SIrange[range-phrase=--,range-units=single]{0.8}{8}{\second}. Finally, comparing $t_{\text{deploy}}$ shows ArgoCD performing better with median latency around \SI{200}{\milli\second} up to $r=50$ and it shoots up linearly to \SI{10}{\second}, converging with the latency of Flux CD and ConfigSync.

In summary, we observed that ArgoCD and FluxCD exhibit more consistent performance compared to ConfigSync. Notably, ArgoCD is slightly quicker in reconciliation and interfacing with the back-end Kubernetes cluster for deployment, compared to FluxCD.

\subsection{Test-case 2A: Multi-App Latency Comparison}

\begin{figure}
    \centering
    \includegraphics[width=1\linewidth]{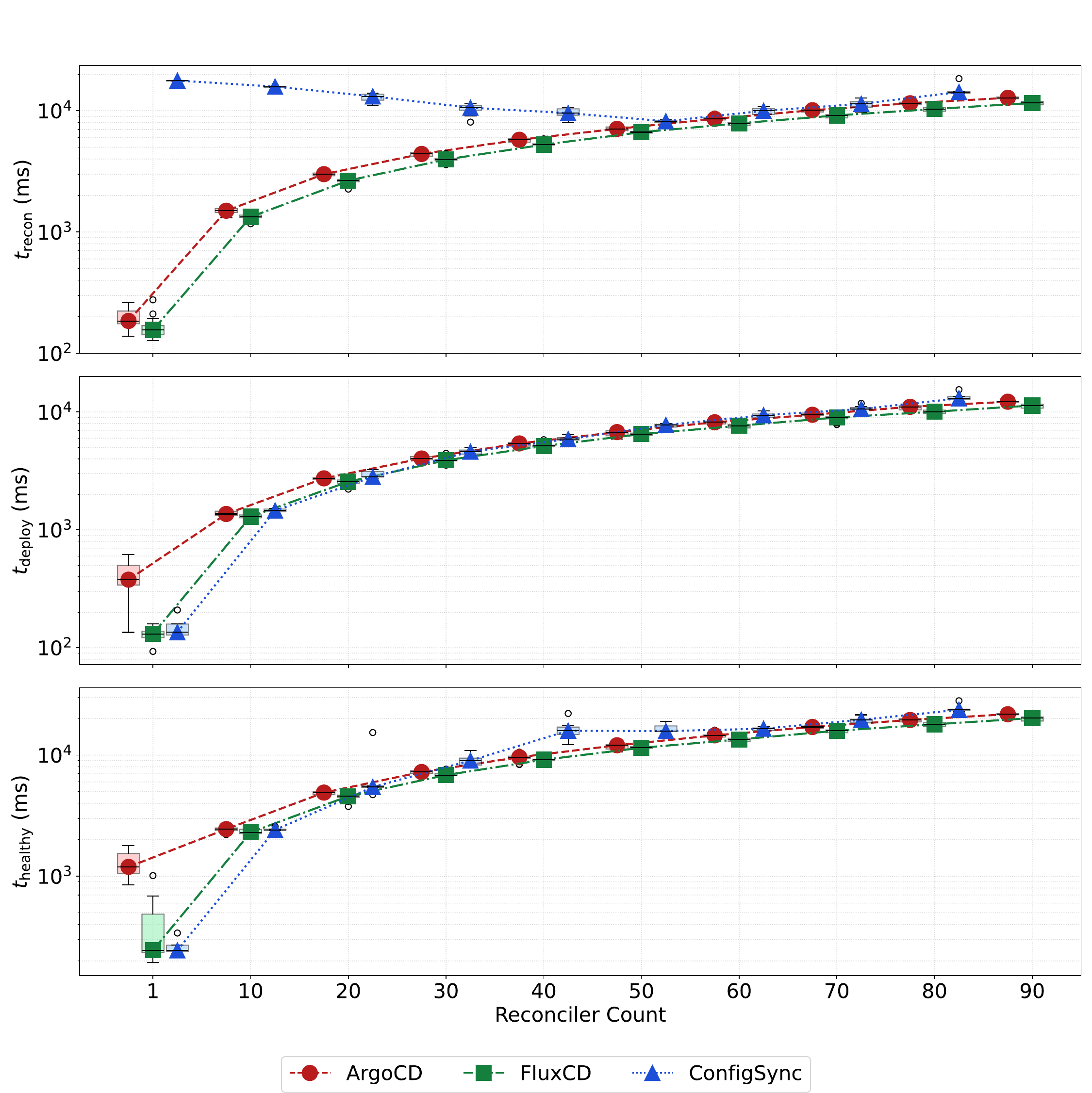}
    \caption{Latency Comparison of GitOps Tools with Scaling Concurrent Multi-App Deployment}
    \label{fig:mult-app-lat}
\end{figure}

Fig. \ref{fig:mult-app-lat} depicts the combined measurements of the experiments. All three reconcilers show an identical trend of $t_{\text{deploy}}$ and $t_{\text{healthy}}$. The $t_{\text{healthy}}$ result is as expected, as it is beyond the scope of the Reconcilers, similar to the case of $t_{\text{push}}$ in the testcase 1; therefore, we omit it in this testcase intentionally. We attribute the linear growth of $t_{\text{deploy}}$ to minor fluctuations resulting from parallelising the deployment of non-scaling applications, which Kubernetes schedulers take advantage of. However, there is a trade-off in resource utilisation, which is revealed in the next sub-section. After conducting the experiments several times, we observed a "V-Shaped" pattern emerging from $t_{\text{recon}}$ of ConfigSync. It starts with a \SI{17.5}{\second} value, gradually converges with that of the other two at $r=50$ to \SI{7.5}{\second}. This is roughly the same as its $t_{\text{recon}}$ for a single-app test case, and follows the same growth trajectory as Argo CD and Flux CD from there. At the time of writing this article, we don't have a clear explanation for this observation; however, we aim to investigate it further in our future work.

In summary, all reconcilers performed equally after concurrent deployment of the application with $r \ge 50$. Argo CD and Flux CD are the most consistent throughout, while ConfigSync shows convergence in cases of a large number of concurrent requests, with less fluctuation compared to Testcase 1.

\subsection{Test-Case 2B: Multi-App Resource Utilisation}
\begin{figure}
    \centering
    \includegraphics[width=1\linewidth]{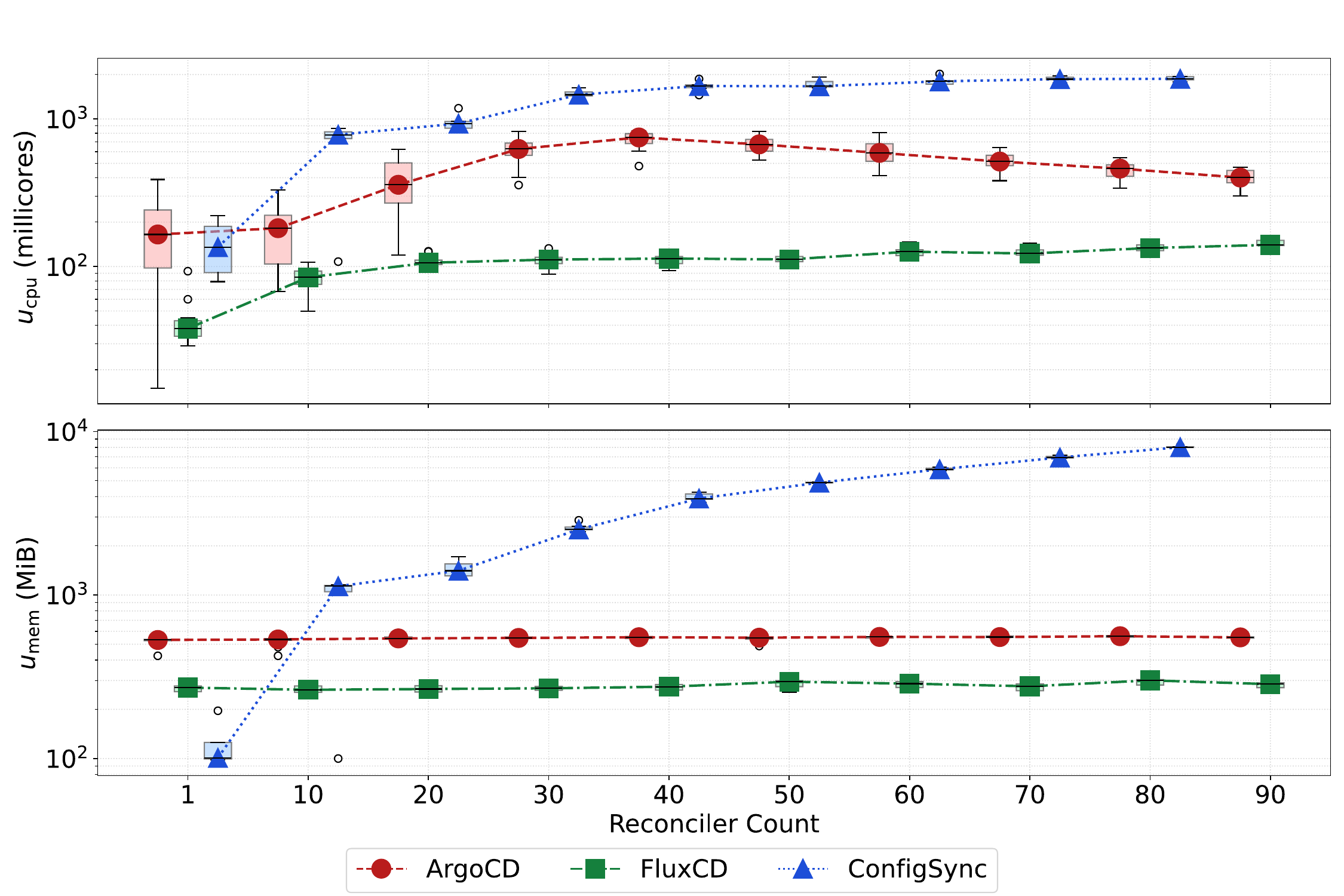}
    \caption{Resource Utilisation Comparison of GitOps Tools with Scaling Concurrent Multi-App Deployment}
    \label{fig:multi-app-util}
\end{figure}

For this investigation (Fig. \ref{fig:multi-app-util}), we measured the $u_{\text{cpu}}$ and $u_{\text{mem}}$ from K8s Container Runtime Interface. The median $u_{\text{cpu}}$ stays within \SIrange[range-phrase=--,range-units=single]{1}{250}{} millicore for Flux CD, \SIrange[range-phrase=--,range-units=single]{150}{750}{} millicore for Argo CD and \SIrange[range-phrase=--,range-units=single]{120}{1900}{} millicore for ConfigSync. The median $u_{\text{mem}}$ remains consistent at \SI{120}{\mega\byte} and \SI{520}{\mega\byte} for Flux CD and Argo CD, respectively, with a significant difference from ConfigSync reaching up to \SI{8}{\giga\byte} at $r=90$. Our investigation suggests that the resource-intensive behaviour of ConfigSync is due to its concurrency handling mechanism. Unlike Flux CD and Argo CD, which run a single reconciler in their respective control plane namespaces (i.e., flux-system and argo, respectively), ConfigSync instantiates individual root-reconciler objects in the config-management-system namespace to bind each application, resulting in a significant overhead as the number of concurrent application requests scales.

In summary, Flux CD is the least resource-intensive, Argo CD has a similar memory footprint to Flux CD with higher CPU consumption, which settles down to a level comparable to Flux CD after $r=40$, and ConfigSync is the most resource-intensive in both CPU and memory consumption.

\subsection{Test-Case 3: Nephio's Intent Processing Latency}
\begin{figure}[t]
    \centering
    \includegraphics[width=1\linewidth]{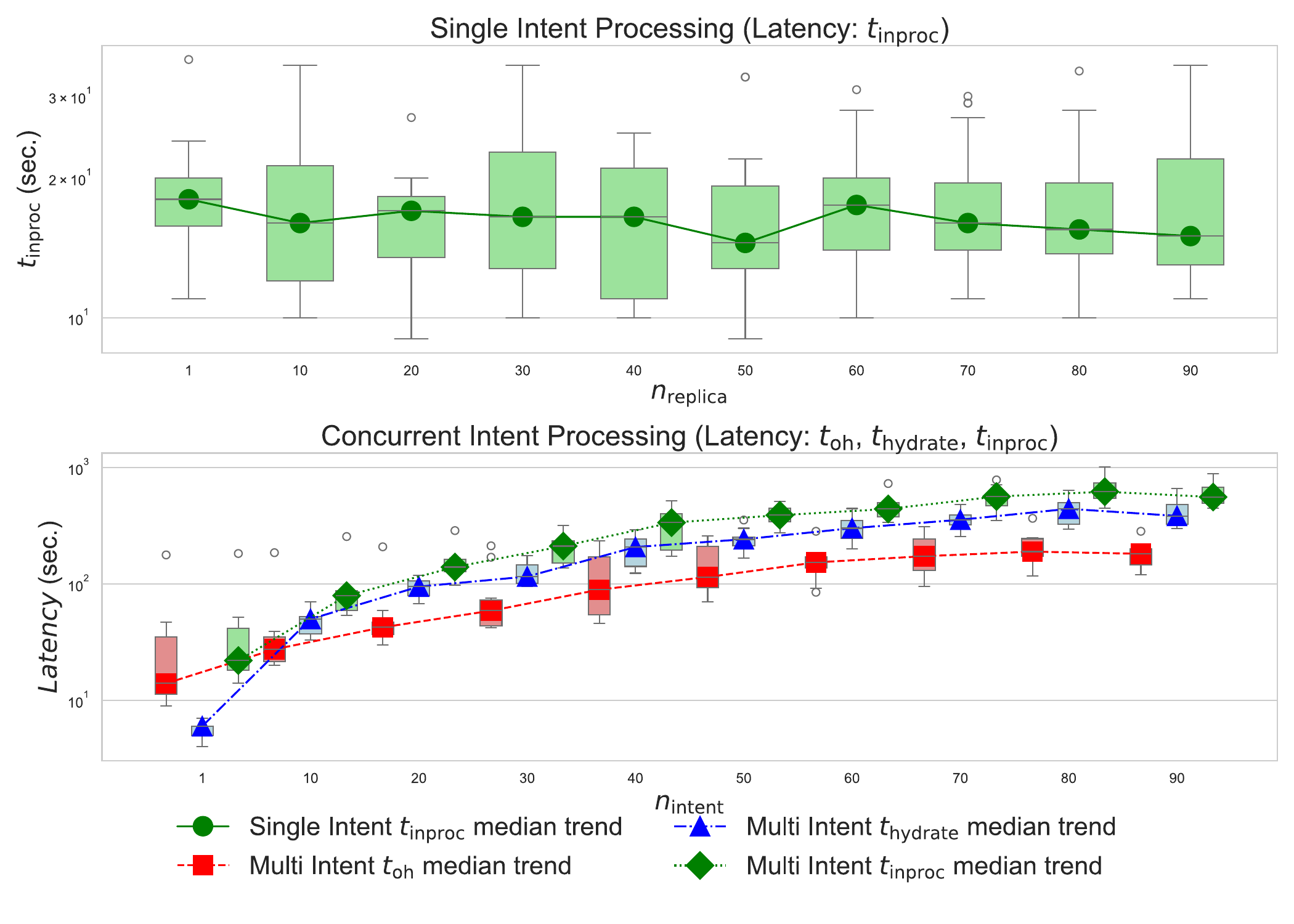}
    \caption{Comparison of Nephio's Intent Processing Latency}
    \label{fig:nephio-latency}
\end{figure}

Fig. \ref{fig:nephio-latency} illustrates the latency profile of Nephio in processing the Intent ($t_{\text{intproc}}$) first, for a single deployment of Intent, scaling by replica; and second; for multiple deployment by scaling the number of Dry Packages and their corresponding PVs both within a range of $[1:90:10]$. The intent processing latency $t_{\text{intproc}} = t_{\text{hydration}} + t_{\text{oh}}$, where $t_\text{{oh}}$ is the latency introduced by the overhead processing including time to bring up the PVs, Webhook establishment between the PVs \& BPR and discovery of Draft Packages in DPR. The $t_{\text{oh}}$ is proportional to the number of Intents. The experimental result shows that Nephio introduces a mean constant $t_{\text{intproc}}$ for Intent deployment, which is \SIrange[range-phrase=--,range-units=single]{17.62}{23.85}{\second} per Intent, including the default reconciliation period configured in Porch and the overhead of instantiating PV. In case of the multi-Intent experiments, we observed the mean $t_{\text{inproc}} = $\SI{11.2}{\second} with its components  $t_{\text{oh}}=$ \SI{6.16}{\second} and $t_{\text{hydrate}}=$ \SI{4.95}{\second}. The mean $t_{\text{inproc}}$ for multi-Intent is slightly lower (by $\sim$ \SI{6}{\second}) compared to that of the single-Intent because K8s instantiates the PVs simultaneously; therefore, on average, the latency due to overhead processing lowers as the number of Intents scales.

\subsection{Summary and Findings}

\begin{table*}[t]
\centering
\caption{Summary statistics of various latency \& utilisation KPIs in single \& multiple Intent deployment scenarios through GitOps tools (Argo CD, Flux CD \& ConfigSync) and IBN Orchestrator (Nephio)}
\renewcommand{\arraystretch}{1.2}
\begin{tabular}{c|cccccccccc}
\cline{1-9}
\multicolumn{1}{|c|}{\cellcolor[HTML]{EFEFEF}} & \multicolumn{8}{c|}{\cellcolor[HTML]{EFEFEF}\textbf{Single   Intent Deployement}} &  &  \\ \cline{2-9}
\multicolumn{1}{|c|}{\cellcolor[HTML]{EFEFEF}} & \multicolumn{2}{c|}{\cellcolor[HTML]{EFEFEF}$t_{\text{push}}$  (sec.)} & \multicolumn{2}{c|}{\cellcolor[HTML]{EFEFEF}$t_{\text{sync}}$ (sec.)} & \multicolumn{2}{c|}{\cellcolor[HTML]{EFEFEF}$t_{\text{recon}}$ (sec.)} & \multicolumn{2}{c|}{\cellcolor[HTML]{EFEFEF}$t_{\text{deploy}}$ (sec.)} &  &  \\ \cline{2-9}
\multicolumn{1}{|c|}{\multirow{-3}{*}{\cellcolor[HTML]{EFEFEF}\textbf{Tools}}} & \multicolumn{1}{c|}{\cellcolor[HTML]{EFEFEF}$\mu$} & \multicolumn{1}{c|}{\cellcolor[HTML]{EFEFEF}$\sigma$} & \multicolumn{1}{c|}{\cellcolor[HTML]{EFEFEF}$\mu$} & \multicolumn{1}{c|}{\cellcolor[HTML]{EFEFEF}$\sigma$} & \multicolumn{1}{c|}{\cellcolor[HTML]{EFEFEF}$\mu$} & \multicolumn{1}{c|}{\cellcolor[HTML]{EFEFEF}$\sigma$} & \multicolumn{1}{c|}{\cellcolor[HTML]{EFEFEF}$\mu$} & \multicolumn{1}{c|}{\cellcolor[HTML]{EFEFEF}$\sigma$} &  &  \\ \cline{1-9}
\multicolumn{1}{|c|}{\cellcolor[HTML]{EFEFEF}\textbf{Argo CD}} & \multicolumn{1}{c|}{$1.05$} & \multicolumn{1}{c|}{$0.01$} & \multicolumn{1}{c|}{$2.83$} & \multicolumn{1}{c|}{$0.37$} & \multicolumn{1}{c|}{$0.01$} & \multicolumn{1}{c|}{$0.01$} & \multicolumn{1}{c|}{$9.07$} & \multicolumn{1}{c|}{$0.04$} &  &  \\ \cline{1-9}
\multicolumn{1}{|c|}{\cellcolor[HTML]{EFEFEF}\textbf{Flux CD}} & \multicolumn{1}{c|}{$1.04$} & \multicolumn{1}{c|}{$0.01$} & \multicolumn{1}{c|}{$2.58$} & \multicolumn{1}{c|}{$0.09$} & \multicolumn{1}{c|}{$0.0056$} & \multicolumn{1}{c|}{$0.0034$} & \multicolumn{1}{c|}{$0.02$} & \multicolumn{1}{c|}{$0.02$} &  &  \\ \cline{1-9}
\multicolumn{1}{|c|}{\cellcolor[HTML]{EFEFEF}\textbf{Config Sync}} & \multicolumn{1}{c|}{$1.01$} & \multicolumn{1}{c|}{$0.02$} & \multicolumn{1}{c|}{$217.53$} & \multicolumn{1}{c|}{$112.15$} & \multicolumn{1}{c|}{$0.03$} & \multicolumn{1}{c|}{$0.05$} & \multicolumn{1}{c|}{$0.11$} & \multicolumn{1}{c|}{$0.01$} &  &  \\ \hline
 & \multicolumn{10}{c|}{\cellcolor[HTML]{EFEFEF}\textbf{Multiple Intent Deployment}} \\ \cline{2-11}
 & \multicolumn{2}{c|}{\cellcolor[HTML]{EFEFEF}$t_{\text{recon}}$ (sec.)} & \multicolumn{2}{c|}{\cellcolor[HTML]{EFEFEF}$t_{\text{deploy}}$ (sec.)} & \multicolumn{2}{c|}{\cellcolor[HTML]{EFEFEF}$t_{\text{healthy}}$ (sec.)} & \multicolumn{2}{c|}{\cellcolor[HTML]{EFEFEF}$u_{\text{cpu}}$ (Milicore)} & \multicolumn{2}{c|}{\cellcolor[HTML]{EFEFEF}$u_{\text{mem}}$ (MB)} \\ \cline{2-11}
\multirow{-3}{*}{} & \multicolumn{1}{c|}{\cellcolor[HTML]{EFEFEF}$\mu$} & \multicolumn{1}{c|}{\cellcolor[HTML]{EFEFEF}$\sigma$} & \multicolumn{1}{c|}{\cellcolor[HTML]{EFEFEF}$\mu$} & \multicolumn{1}{c|}{\cellcolor[HTML]{EFEFEF}$\sigma$} & \multicolumn{1}{c|}{\cellcolor[HTML]{EFEFEF}$\mu$} & \multicolumn{1}{c|}{\cellcolor[HTML]{EFEFEF}$\sigma$} & \multicolumn{1}{c|}{\cellcolor[HTML]{EFEFEF}$\mu$} & \multicolumn{1}{c|}{\cellcolor[HTML]{EFEFEF}$\sigma$} & \multicolumn{1}{c|}{\cellcolor[HTML]{EFEFEF}$\mu$} & \multicolumn{1}{c|}{\cellcolor[HTML]{EFEFEF}$\sigma$} \\ \hline
\multicolumn{1}{|c|}{\cellcolor[HTML]{EFEFEF}\textbf{Argo CD}} & \multicolumn{1}{c|}{$0.14$} & \multicolumn{1}{c|}{$0.01$} & \multicolumn{1}{c|}{$0.14$} & \multicolumn{1}{c|}{$0.08$} & \multicolumn{1}{c|}{$0.24$} & \multicolumn{1}{c|}{$0.002$} & \multicolumn{1}{c|}{$13.46$} & \multicolumn{1}{c|}{$6.26$} & \multicolumn{1}{c|}{$10.1$} & \multicolumn{1}{c|}{$7.11$} \\ \hline
\multicolumn{1}{|c|}{\cellcolor[HTML]{EFEFEF}\textbf{Flux CD}} & \multicolumn{1}{c|}{$0.13$} & \multicolumn{1}{c|}{$0.01$} & \multicolumn{1}{c|}{$0.13$} & \multicolumn{1}{c|}{$0.002$} & \multicolumn{1}{c|}{$0.23$} & \multicolumn{1}{c|}{$0.01$} & \multicolumn{1}{c|}{$2.24$} & \multicolumn{1}{c|}{$2.28$} & \multicolumn{1}{c|}{$5.34$} & \multicolumn{1}{c|}{$3.4$} \\ \hline
\multicolumn{1}{|c|}{\cellcolor[HTML]{EFEFEF}\textbf{Config Sync}} & \multicolumn{1}{c|}{$0.18$} & \multicolumn{1}{c|}{$0.18$} & \multicolumn{1}{c|}{$0.15$} & \multicolumn{1}{c|}{$0.01$} & \multicolumn{1}{c|}{$0.28$} & \multicolumn{1}{c|}{$0.05$} & \multicolumn{1}{c|}{$33.31$} & \multicolumn{1}{c|}{$9.97$} & \multicolumn{1}{c|}{$98.47$} & \multicolumn{1}{c|}{$1.56$} \\ \hline
\rowcolor[HTML]{EFEFEF}
\multicolumn{1}{|c|}{\cellcolor[HTML]{EFEFEF}} & \multicolumn{4}{c|}{\cellcolor[HTML]{EFEFEF}\textbf{Single Intent Deployment}} & \multicolumn{6}{c|}{\cellcolor[HTML]{EFEFEF}\textbf{Multiple Intent Deployment}} \\ \cline{2-11}
\rowcolor[HTML]{EFEFEF}
\multicolumn{1}{|c|}{\cellcolor[HTML]{EFEFEF}} & \multicolumn{4}{c|}{\cellcolor[HTML]{EFEFEF}$t_{\text{inproc}}$ (sec.)} & \multicolumn{2}{c|}{\cellcolor[HTML]{EFEFEF}$t_{\text{inproc}}$ (sec.)} & \multicolumn{2}{c|}{\cellcolor[HTML]{EFEFEF}$t_{\text{hydrate}}$ (sec.)} & \multicolumn{2}{c|}{\cellcolor[HTML]{EFEFEF}$t_{\text{oh}}$ (sec.)} \\ \cline{2-11}
\rowcolor[HTML]{EFEFEF}
\multicolumn{1}{|c|}{\cellcolor[HTML]{EFEFEF}} & \multicolumn{2}{c|}{\cellcolor[HTML]{EFEFEF}$\mu$} & \multicolumn{2}{c|}{\cellcolor[HTML]{EFEFEF}$\sigma$} & \multicolumn{1}{c|}{\cellcolor[HTML]{EFEFEF}$\mu$} & \multicolumn{1}{c|}{\cellcolor[HTML]{EFEFEF}$\sigma$} & \multicolumn{1}{c|}{\cellcolor[HTML]{EFEFEF}$\mu$} & \multicolumn{1}{c|}{\cellcolor[HTML]{EFEFEF}$\sigma$} & \multicolumn{1}{c|}{\cellcolor[HTML]{EFEFEF}$\mu$} & \multicolumn{1}{c|}{\cellcolor[HTML]{EFEFEF}$\sigma$} \\ \cline{2-11}
\multicolumn{1}{|c|}{\multirow{-4}{*}{\cellcolor[HTML]{EFEFEF}\textbf{Nephio}}} & \multicolumn{2}{c|}{$17.9$} & \multicolumn{2}{c|}{$0.91$} & \multicolumn{1}{c|}{$7.79$} & \multicolumn{1}{c|}{$0.34$} & \multicolumn{1}{c|}{$4.97$} & \multicolumn{1}{c|}{$0.42$} & \multicolumn{1}{c|}{$2.73$} & \multicolumn{1}{c|}{$0.13$} \\ \hline
\end{tabular}
\label{tab:stat-summary}
\end{table*}

Table \ref{tab:stat-summary} summarises the performance evaluation against all KPIs $K_{attr}$ from our experiments. Our findings show the following. First, the $t_{\text{sync}}$ \& $t_{\text{recon}}$ of ConfigSync is higher and less deterministic compared to that of Argo CD and Flux CD. Second, Flux CD is more compute-intensive but faster than Argo CD. Third, Nephio's overall Intent processing latency with default Porch settings is almost constant per intent, and it scales linearly with the number of Intents.

To summarise each scenario, we processed the dataset as follows. First, we standardised the dataset concerning the scaling variable, i.e., the number of replicas for single intent and the number of apps for multi-intent use cases. Thereafter, we filter out the outliers by aggregating them based on their median value. For standard deviation $\sigma$, we computed the sample $\sigma$ after removing statistical outliers using the Interquartile Range (IQR). Specifically, any data point lying outside the interval $[Q_1 - 1.5 \cdot \text{IQR},\ Q_3 + 1.5 \cdot \text{IQR}]$ was excluded. The $\sigma$ was then calculated over the remaining values using the unbiased estimator.

\section{Conclusion \& Future Scope}
\label{sec:conclusion}
This work investigates the effect of integrating three GitOps operators (i.e., Argo CD, Flux CD, and ConfigSync) and an IBN orchestrator, Nephio, to enhance the E2E Intent Deployment performance of our bespoke orchestration platform, CAMINO. It describes a benchmarking pipeline with a methodology to collect defined KPIs that summarise the performance evaluation through latency and resource utilisation in single and multi-intent deployment scenarios, supported by statistical analysis of experimental results.

In this benchmarking setup, we utilised a homogeneous intent deployment model by deploying a simple Nginx application to measure all KPIs; that said, we have established a configurable testbed infrastructure. Hence, as an extension, we shall leverage it to conduct benchmarking in a heterogeneous setup with a variety of standard Network Functions (e.g., 5G Core). In addition, we aim to advance this framework as a tool to generate large-scale datasets to train machine-learning models, enabling the prediction of anticipatory latency and resource consumption in an intent-deployment scenario. AI-Native IBN for a 6G communication system would leverage such predictions for optimising proactive resource allocation.

\section{Acknowledgement}
This work is funded in part by REASON+, a UK Government-funded project under the Future Open Networks Research Challenge (FONRC), sponsored by the Department of Science, Innovation, and Technology (DSIT), and in part by Innovate UK funding under the CelticNext SUSTAINET-guarDian project.

\bibliographystyle{IEEEtran}
\bibliography{ref}

\end{document}